\documentclass[aps,prl,amsfonts,amssymb,epsfig,twocolumn,superscriptaddress,showpacs]{revtex4}

\usepackage{graphics,graphicx}
\usepackage{amsmath,bbm}
\usepackage{amstext}
\usepackage{amssymb}
\usepackage{color}

\def\beq{\begin{equation}}
\def\eeq{\end{equation}}
\def\bea{\begin{eqnarray}}
\def\eea{\end{eqnarray}}
\providecommand{\openone}{\leavevmode\hbox{\small1\kern-3.8pt\normalsize1}}

\begin{document}

\title{Quantum MERA Channels}

\author{V. Giovannetti}
\affiliation {NEST CNR-INFM \& Scuola Normale Superiore, Piazza
dei Cavalieri 7, I-56126 Pisa, Italy}
\author{S. Montangero}
\affiliation {NEST CNR-INFM \& Scuola Normale Superiore, Piazza
dei Cavalieri 7, I-56126 Pisa, Italy}
\author{Rosario Fazio}
\affiliation {NEST CNR-INFM \& Scuola Normale Superiore, Piazza
dei Cavalieri 7, I-56126 Pisa, Italy}
\affiliation {International School for Advanced Studies (SISSA),
 Via Beirut 2-4, I-34014 Trieste, Italy}

\date{\today}

\begin{abstract}
Tensor networks representations  of 
many-body quantum systems can be described in terms of quantum channels.
We focus on channels associated with the 
Multi-scale Entanglement Renormalization Ansatz (MERA) tensor 
network that has been recently introduced to efficiently describe critical systems.
Our approach allows us to compute the MERA correspondent to the
thermodynamic limit of a critical system introducing a transfer matrix formalism, 
and to relate the system critical exponents to the convergence rates of the associated channels.  
\end{abstract}


\maketitle
Understanding the properties of strongly interacting many-body quantum system is 
central in many areas of physics. Whenever it is hard to device reliable 
analytical approaches, as in many situations of experimental relevance,  
ingenious numerical methods are necessary  to grasp the essential 
properties of these systems. Our ability of simulating  them is based on the possibility 
to find an efficient description of their ground state.
This is the case, for example, of White's Density Matrix 
Renormalization Group~\cite{DMRG1} which can be recasted in terms of Matrix Product States
(MPS)~\cite{FNW,OR,VIDAL1,VPC,rev}.   
Such representations  are characterized by a 
simple  tensor  decomposition of the many-body wave-function which allows one  {\em i)} 
to efficiently compute all the  relevant observables of the system (e.g. energy, 
local observables,  and correlation functions), and {\em ii)} to reduce the 
effective number of parameters over which the numerical optimization  needs to 
be performed. MPS fulfill these requirements and  can be used to describe faithfully
the ground states of not critical, short range one-dimensional many-body Hamiltonians
at zero-temperature. However MPS typically fail to provide an accurate
description in other relevant situations, i.e. when the system is
critical, in higher physical dimensions or if the model possesses
long-range couplings. Several proposals have been put forward to
overcome this problem.  
Projected Entangled Pair States (PEPS)~\cite{peps} generalize MPS in dimensions 
higher than one.  Weighted graph states~\cite{wgs} can deal with long-range correlations. 
In this Letter we focus on a solution recently proposed by 
Vidal~\cite{mera} who introduced a tensor structure based on the so called
Multiscale Entanglement Renormalization Ansatz (MERA). The MERA tensor network
satisfies both the constraints {\em i)} and {\em ii)} and accommodates 
the scale invariance typical of critical systems~\cite{meraalg,meraapp}.
The relevance of this approach might 
represent a major breakthrough in our simulation capabilities~\cite{2DMERA} 
and motivates an intensive study of the MERA~\cite{vidallong,DEO}. 

Here we point out a previously unnoticed connection between the MERA 
and the theory of completely positive quantum maps~\cite{BENGZY}
establishing a  link between two important areas of quantum information science.
This allows us to introduce a transfer matrix formalism
 in the same spirit as it has been done for MPS~\cite{VPC,OR}, providing 
new tools to compute physical observables using MERA. 
The main outcomes of our work 
are i) a  method for determining the properties of critical many-body systems in the thermodynamic 
limit and ii) a {\em connection between the critical exponents} governing the decay of 
correlation functions and {\em the eigenvalues of the MERA transfer matrix}.
As a consequence this yields a full characterization of the asymptotic properties of 
one-dimensional critical systems. 
The paper is organized as follows:
after a brief review of the MERA, we show how the
 local expectation values and correlations functions can be  casted in terms 
of concatenated quantum channels. Using general properties of {\em mixing} 
quantum channels~\cite{NJP,RAG,GOHM,TDV} we then provide a way for expressing the 
thermodynamic limit of these quantities. 

\begin{figure}[t!]
\begin{center}
\includegraphics[scale=0.45]{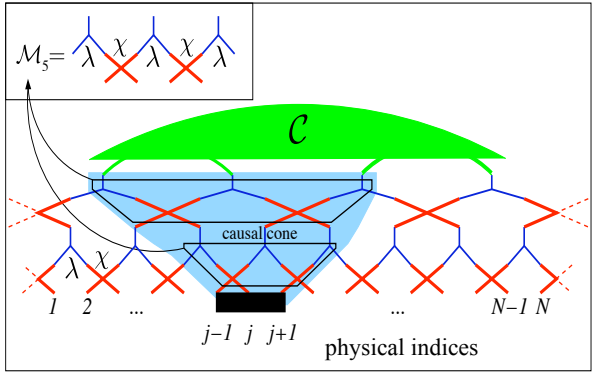}
\caption{Representation of a MERA decomposition of the
tensor ${\cal T}$ associated with a many-body state $|\psi\rangle$ 
 for $N=16$ qudits (a dotted link emerging from the left side of the graph re-enter the figure as the corresponding dotted
 link on the right). The light blue region represents the causal cone~\cite{mera}
 associated with the local operator
 $\hat{\Theta}_j$ (black rectangle). It is composed  by contraction of tensors ${\cal M}_5$  (shown
 in the inset). 
} \label{fig1}
\end{center}
\end{figure}

\paragraph{The MERA tensor network:--} Consider a many-body quantum system composed by
$N =2^n$ sites of dimension $d$ (qudits). 
Any pure state can be expressed as 
$|\psi\rangle = \sum 
 \; {\cal T}_{\ell_1,\ell_2,\cdots, \ell_N} 
 | \xi_{\ell_1} ,\xi_{\ell_2} , \cdots ,  \xi_{\ell_N}\rangle$,
 where for $j\in\{1,\cdots,N\}$ and $\ell \in\{ 1, \cdots, d\}$ the vectors
 $|\xi_{\ell}\rangle_j\in{\cal H}_d$ form the basis of the $j$-th $d$-dimensional
 system component.
The MERA  representation~\cite{mera} of  $|\psi\rangle$ assumes a specific
tensor decomposition of ${\cal T}$ described in Fig.~\ref{fig1}. 
Here the links emerging from the lowest part of the graph represent the $N$  {\em physical indices}
of ${\cal T}$.
The nodes of the graph instead
 represent  tensors. They are divided in three groups: the 
 type-$
 \mbox{\tiny{$\left(\begin{array}{c} 2 \\2
 \end{array}\right)$}}$ {\em disentangler} tensors $\chi$    of
 elements ${\chi}^{u_1,u_2}_{\ell_1,\ell_2}$ 
represented by the red
 Xs;  
 the type-$
 \mbox{\tiny{$\left(\begin{array}{c} 1 \\2
 \end{array}\right)$}}$
 tensors ${\lambda}$  of elements $\lambda^{u_1}_{\ell_1,\ell_2}$ 
represented by the blue inverted Ys;
 and  the type-$
 \mbox{\tiny{$\left(\begin{array}{c} 0 \\ 4
 \end{array}\right)$}}$
tensor $\cal C$ of elements
${\cal C}_{\ell_1,\ell_2,\ell_3,\ell_4}$, represented by the green semi-circle.
As shown in Fig.~\ref{fig1} the  $\chi$'s, the $\lambda$'s are coupled together to form
a triangular structure 
with  ${\cal C}$ as the closing element of the top: any two joined legs from any two distinct 
nodes indicate saturation of 
the associated indices. Consequently, the tensor ${\cal T}$ associated with the $N$ 
qudit state $|\psi\rangle$ is  written as a
network of $O(N)$ tensors
organized in 
$O(\log_2 N)$ 
different levels composed by one layer of ${\chi}$ tensors connected with one layer of $\lambda$ tensors.
In a generic MERA the $\chi$'s and the $\lambda$'s  differ
from node to node but  the dimensions of their indices 
are upper-bounded by a fixed constant (for easy of notation we
omit the labels expressing the position of the tensors in the network).

What makes the MERA decomposition a convenient one is the 
assumption that the ${\chi}$'s and $\lambda$'s satisfy special 
contraction rules. Specifically,  for each $\chi$ and $\lambda$ let us define 
its adjoint $\bar{\chi}$ and $\bar{\lambda}$ as the 
tensors of elements $\bar{\chi}_{u_1,u_2}^{ \ell_1, \ell_2} =({\chi}_{
  \ell_1, \ell_2}^{u_1,u_2})^*$
and $\bar{\lambda}^{u_1,u_2}_{ \ell_1} =({\lambda}^{ \ell_1 }_{ u_1,
  u_2})^*$.
  With this definition the
MERA contraction rules are 
$
\bar{\chi}^{\bullet, \diamond}_{ \ell_1, \ell_2} \; {\chi}_{ \bullet,\diamond}^{u_1,u_2}= 
{\chi}^{ \bullet, \diamond}_{u_1,u_2} \; \bar{\chi}_{\bullet, \diamond}^{\ell_1, \ell_2}
=\delta_{\ell_1,u_1}\delta_{\ell_2,u_2} 
$
and
$\bar{\lambda}^{\bullet , \diamond}_{\ell_1} \;
\lambda_{\bullet , \diamond}^{u_1}
= \delta_{\ell_1,u_1}$,
with $\delta$ being the Kronecker delta
 and where the typographic symbols $\bullet$ and $\diamond$
indicate summation over the corresponding index.
Under these conditions the expectation values of local observables 
on $|\psi\rangle$ requires only to evaluate $O(\log_2 N)$ non trivial 
tensor contractions~\cite{mera}. 

\paragraph{Local observables and quantum channels:-} We first show how the average 
of local observables can be related to the study of concatenated quantum channels.  
Given   an operator $\hat{\Theta}_j$ which acts not trivially  on no more than three consecutive 
qudits (say the  $(j-1)$th, $j$th and $(j+1)$th), the quantity $\langle
\hat{\Theta}_j\rangle \equiv  
\langle \psi| \hat{\Theta}_j|\psi\rangle$ requires to perform contractions only 
over the $\chi$'s and $\lambda$'s 
 belonging to the {\em causal cone}~\cite{mera} 
of the triple $j-1$, $j$ and $j+1$.
A compact  expression 
is obtained by grouping these tensors
in compounds composed  by 2 $\chi$'s and by 3 $\lambda$'s (see  inset of 
  Fig.~\ref{fig1}). This forms  $m=\log_2(N/4)$ non necessarily identical 
type-$\mbox{\tiny{$\left(\begin{array}{c} 3 \\ 6
 \end{array}\right)$}}$   tensors 
${\cal M}_5 \equiv \lambda \chi \lambda \chi\lambda$
where the products $\lambda \chi$  and  $ \chi\lambda$ are defined by 
$[\lambda \chi]^{u_1, u_2}_{\ell_1,\ell_2, \ell_3} \equiv
\lambda^{u_1}_{\ell_1,  \bullet} \;
\chi^{\bullet, u_2 }_{\ell_2, \ell_3}$, and 
$[\chi \lambda]^{u_1, u_2}_{\ell_1,\ell_2 ,\ell_3} \equiv
\chi^{u_1, \bullet}_{\ell_1, \ell_2 } \;
\lambda^{u_2}_{\bullet , \ell_3}$.
For each one of the $m$ tensors ${\cal M}_5$ 
we can then introduce two families of operators 
$\{\hat{L}_{r}\}_{r}$ and $\{ \hat{R}_{r}\}_{r}$ 
acting on the
Hilbert space ${\cal H}_d^{\otimes 3}$ and labeled through the composed index
${r} \equiv ( r_1, r_2,r_3)$ with $r_{1,2,3}$ being $d$-dimensional~\cite{NOTA}. In the computational
basis they are 
defined by the matrices $\langle \xi_{u_1} , \xi_{u_2},  \xi_{u_3} |\hat{L}_{r} |
 \xi_{\ell_1} , \xi_{\ell_2} , \xi_{\ell_3}\rangle$ and 
 $\langle \xi_{u_1} , \xi_{u_2},  \xi_{u_3} |\hat{R}_{r} |
 \xi_{\ell_1} , \xi_{\ell_2} , \xi_{\ell_3}\rangle$ of elements 
$[{\cal M}_5]^{u_1,u_2,u_3}_{r_1,
  \ell_1,\ell_2,\ell_3,r_2,r_3}$ and 
$ [{\cal M}_5]^{u_1,u_2,u_3}_{
 r_1, r_2,\ell_1,\ell_2,\ell_3,r_3 }$
respectively.
They are related through a reshuffling of the
input and output qudits, i.e. $\hat{R}_{r}= \Pi( \hat{L}_{r}) \equiv
\hat{P} \hat{L}_{r} \hat{P}^\dag$,
where $\hat{P} = \hat{P}^\dag$ is the unitary transformation which exchanges 
the first and the third qudit.
Most importantly, according to the contraction rules defined previously, they  
satisfy the  normalization 
conditions $\sum_{r} \hat{L}_{r}\hat{L}_{r}^\dag = \hat{I}^{\otimes 3} = \sum_{r} 
\hat{R}_{r} \hat{R}_{r}^\dag$,
with $\hat{I}$ being the identity operator of ${\cal H}_d$.
This implies that $\{ \hat{L}_{r}\}_{r}$ 
can be used to define
a completely positive, unital, not necessarily trace preserving super-operators 
$\Phi_H^{(L)}$~\cite{BENGZY},
which transforms the linear operator $\hat{\Theta}$ of ${H}_d^{\otimes 3}$ into 
$\Phi_H^{(L)}(\hat{\Theta}) = \sum_{r} \hat{L}_{r} \hat{\Theta} \hat{L}_{r}^\dag$.
Analogously $\{ \hat{R}_{r}\}_{r}$   defines the map $\Phi_H^{(R)}$ 
which is related with $\Phi_H^{(L)}$ through the identity
$\Phi_H^{(R)} = \Pi \circ \Phi_H^{(L)} \circ \Pi$,
where "$\circ$" indicates the  composition of super-operators.
We also introduce the  vector of 
 ${\cal H}_d^{\otimes 4}$,
$|hat\rangle \equiv  \sum 
{\cal C}_{\ell_1,\ell_2,\ell_3,\ell_4}  \; |\xi_{\ell_1},
 \xi_{\ell_2},\xi_{\ell_3},\xi_{\ell_4}\rangle$,  
which without loss of generality is assumed to be  normalized, and define  $\hat{\rho}_C$ the
three sites reduced density matrix  obtained by tracing $|hat\rangle\langle hat|$ over one of the $4$ qudits.
With these definitions  one can finally write the expectation value  of $\hat{\Theta}_j$ as 
$\langle \hat{\Theta}_j \rangle = \mbox{Tr} [ \hat{\rho}_C \; \hat{B}^{(m)}_{j}]$,
with 
$\hat{B}_j^{(m)}  \equiv{\Phi_H^{(m)} \circ \cdots  \circ  \Phi_H^{(1)}} (\hat{\Theta}_j)$,
and  
where (enumerating from the lower MERA level of Fig.~\ref{fig1})  $\Phi_H^{(k)}$ is either the map $\Phi_H^{(L)}$ or $\Phi_H^{(R)}$ associated with
the $k$-th tensor ${\cal M}_5$ of the causal cone 
(which one depends upon
 $N$ and $j$).
The operator $\hat{B}_j^{(m)}$ is thus obtained by applying to the observable $\hat{\Theta}_j$ 
a sequence of $m$  super-operators associated to the MERA causal cone.  
We can then  write
 \begin{eqnarray}\label{imp}
\langle \hat{\Theta}_j \rangle = 
   \mbox{Tr} [ \Phi^{(1\leftarrow m)}  (\hat{\rho}_C) \;   \hat{\Theta}_j]  \;,
\end{eqnarray}
where 
$\Phi^{(1\leftarrow m)}  \equiv 
{\Phi^{(1)} \circ \cdots  \circ  \Phi^{(m)}}$, 
with  $\Phi^{(k)}$  being the super-operator $\Phi_H^{(k)}$
in Schr\"{o}dinger picture. 
 By construction the $\Phi^{(k)}$ (and hence $ \Phi^{(1\leftarrow m)}$)
 are  
Completely Positive, Trace Preserving (CPT) maps, i.e. quantum
channels 
with Kraus operators~\cite{BENGZY} defined by either the set  $\{ \hat{L}_{r}^\dag\}_{r}$ or $\{ \hat{R}_{r}^\dag\}_{r}$.
Equation~(\ref{imp})  establishes a formal equivalence between the MERA tensor network and the successive application of
a family of CPT maps (the QuMERA family).
Since it  holds for {\em all}  the local observable $\hat{\Theta}_j$ 
this implies that  $ \Phi^{(1\leftarrow m)}  (\hat{\rho}_C)$ coincides with the reduced density matrix $\hat{\rho}_j$ 
of the input state $|\psi\rangle$ associated with  the qudits $j-1$, $j$ and $j+1$, i.e. 
\begin{eqnarray}
\Phi^{(1\leftarrow m)} (\hat{\rho}_C) 
= \hat{\rho}_j \label{NNN}\;.
\end{eqnarray}
\paragraph{Critical systems in the thermodynamic limit: --}
Consider now a family of MERA states $|\psi\rangle$ which 
describes the ground state  of a  
many-body Hamiltonian $\hat{H}$ at criticality. We are specifically interested in the thermodynamic limit of infinitely many sites
(i.e. $N \rightarrow \infty$).
According to the above derivation,  in the limit of large $m$  Eq.~(\ref{NNN}) converges toward
the reduced density matrix  $\hat{\rho}_T$ 
of three  consecutive qubits of the systems  
$\lim_{m \rightarrow \infty}  \Phi^{(1\leftarrow m)}
 (\hat{\rho}_C) = \hat{\rho}_T$.
Of course the above limit should not depend upon the particular causal cone "trajectory" 
one chooses to follow  (the system is translational invariant). Without
loss of generality we can thus pick the one associated with the central sites  of the MERA, i.e. 
the one associated with the causal cone of $N/2$-th qudit. This allows us to identify all the
$\Phi^{(k)}$ of $\Phi^{(1\leftarrow m)}$ with maps of the form
$\Phi^{(R)}$. 
A further simplification arises by enforcing  the scale invariance property of the system.
This can be done for instance by assuming that all the tensors $\chi$'s
and $\lambda$'s of the MERA to be identical~\cite{mera}
and by requiring $\Phi^{(L)} =\Phi^{(R)}=\Phi$~\cite{ReqL}.
With this assumption all  the sequence $\Phi^{(1\leftarrow m)}$
can now be written as a composition of $m$ identical quantum channels, i.e. 
\begin{eqnarray}
\Phi^{(1\leftarrow m)} = 
\Phi \circ \cdots \circ \Phi 
=[ \Phi]^{m} \;. \label{Nseq}
\end{eqnarray}
By general results on quantum channels  the vast majority of 
CPT maps are known to be
 {\em mixing} (or {\em relaxing})~\cite{NJP,GOHM,RAG,TDV}. 
This means that for a generic choice of $\Phi$,  in the limit 
$m\rightarrow \infty$ the transformation~(\ref{Nseq}) will send
 all input states
into a unique fix point
 identified as the unique eigen-operator of $\Phi$
associated with its largest eigenvalue.
 This property allows us to identify the thermodynamic limit
$\hat{\rho}_T$ of the reduced density matrix $\hat{\rho}_j$ with such an eigenstate. 
As in the case of MPS~\cite{FNW,WOVC}, we can now provide 
a simplified expression for the thermodynamic limit of any local observable 
$\hat{\Theta}$ of scale invariant MERAs.
A convenient way to express this is obtained by moving in Liouville space~\cite{BENGZY,ROYER}.
By doing so we can write
\begin{eqnarray}
\langle \hat{\Theta} \rangle_T  = \lim_{m\rightarrow \infty}  \langle\langle \hat{\Theta} | (\hat{E}_\Phi)^m | \hat{\rho}_C\rangle\rangle 
=\langle\langle \hat{\Theta} | \hat{\rho}_T\rangle\rangle \;, \label{newequ}
 \end{eqnarray}
where the vectors $|\hat{A}\rangle\rangle$ are  the Liouville representations of the operators $\hat{A}$,
while  $\hat{E}_\Phi$ is the Liouville operator associated to the map $\Phi$, i.e. 
\begin{eqnarray}
\hat{E}_\Phi \equiv 
\sum_r \hat{R}_r \otimes \hat{R}_r^{*} \;, \label{TRANSFER}
\end{eqnarray}
(here $\hat{R}_r^*$ is the complex conjugate
of $\hat{R}_r$ evaluated with respect
to the canonical basis).
 Within this formalism the vector $|\hat{\rho}_T\rangle\rangle$  which describes the fix point $\hat{\rho}_T$ of the mixing map $\Phi$
 is also found as the the (unique) eigenvector  correspondent to the unitary eigenvalue 
 of $\hat{E}_\Phi$.
Because of the close similarities  between $\hat{E}_\Phi$  
and the MPS transfer matrix~\cite{WOVC,FNW,VPC,OR}
we dubbed the former the {\em transfer operator} of the MERA.
The existence and uniqueness of the fix point is given by the physical
assumption that the thermodynamical limit of a physical system 
exists and it is unique.

\begin{figure}[t!]
\begin{center}
\includegraphics[scale=0.19]{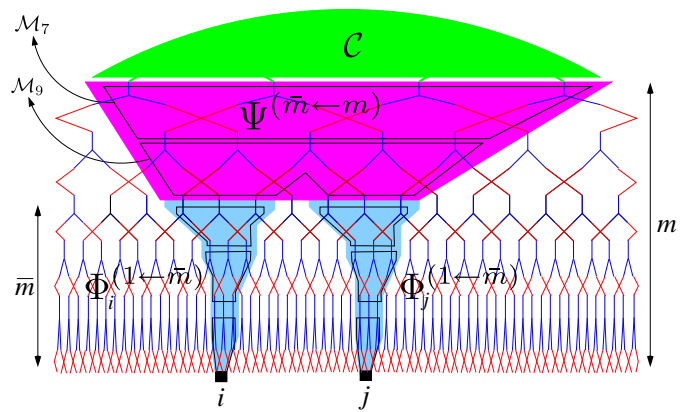}
\caption{Causal cone associated with the two point correlation function
$\langle\hat{\Theta}_i \hat{\Theta}_j\rangle$. The light blue regions
generate the tensor product channel $\Phi^{(1 \leftarrow \bar{m})} \otimes \Phi^{(1 \leftarrow \bar{m})}$.
The magenta region instead corresponds to the CPT map $\Psi^{(\bar{m} \leftarrow {m})}$  
which, acting on $|hat\rangle$  produces
the 6 qudits state $\hat{\rho}^{(\bar{m})}_{ij}$.   
} \label{fig2}
\end{center}
\end{figure}

\paragraph{Correlation functions:--} The computation
of the long range correlation functions of $|\psi\rangle$ has also a
clear interpretation in terms of CPT maps and
the thermodynamic limit can be computed along the same
lines presented for local observables. Here we specialize in the  two
point correlation functions   as the generalization is straightforward. 
Consider then the expectation  value  $\langle \hat{\Theta}_i  \hat{\Theta}_j \rangle$
with $\hat{\Theta}_i$ and $\hat{\Theta}_j$ being two local operators 
acting on (say) the $i$-th and  $j$-th qudit respectively.
In this case the causal cone is formed by two single sites causal cones which
intercept at the MERA level  
$\bar{m} = \mbox{int}[ \log_2(i-j)] -1$,
(see Fig.~\ref{fig2}). The resulting expectation values can then
be written as
\begin{eqnarray}
\langle \hat{\Theta}_i \hat{\Theta}_j \rangle =
  \mbox{Tr} [ (\Phi^{(1 \leftarrow \bar{m})}_i \otimes \Phi^{(1 \leftarrow \bar{m})}_j)( 
 \hat{\rho}_{ij}^{(\bar{m})}) \; ( \hat{\Theta}_i\otimes  \hat{\Theta}_j )] , 
 \label{faf1}
\end{eqnarray}
where 
$\Phi^{(1 \leftarrow \bar{m})}_{i,j}$ 
are the two CPT
maps of the two single-site causal cones associated with the sites $i$ and $j$ respectively
 (light blue regions of Fig.~\ref{fig2}).
The operator $\hat{\rho}_{ij}^{(\bar{m})}$ instead is a $6$ qudits state
associated with the last $m-\bar{m}$ levels of the MERA.
 It is obtained from the $4$ qubit state $|{hat}\rangle$ through the application of a quantum 
channel $\Psi^{(\bar{m} \leftarrow {m})}$ which, similarly to $\Phi_{i,j}^{(1 \leftarrow \bar{m})}$, 
 originates from a proper 
  concatenation of CPT maps associated with ${\cal M}_5$  or with the type-$
 \mbox{\tiny{$\left(\begin{array}{c} 5 \\10
 \end{array}\right)$}}$ and type-$
 \mbox{\tiny{$\left(\begin{array}{c} 4 \\8
 \end{array}\right)$}}$ tensors ${\cal M}_9\equiv \lambda \chi  \lambda \chi  \lambda \chi  
\lambda \chi  \lambda$ and 
 ${\cal M}_7\equiv \lambda \chi  \lambda \chi  \lambda \chi  \lambda$.
 Since this applies to all the two sites observable, we can then conclude that 
 $(\Phi^{(1\leftarrow \bar{m})}_i\otimes \Phi^{(1\leftarrow \bar{m})}_j)(\hat{\rho}_{ij}^{(\bar{m})})$ 
must coincide with the reduced density matrix $\hat{\rho}_{ij}$ of 
$|hat\rangle$ associated with the sites $i$ and $j$.

Let us focus then on the 
thermodynamic limit of the correlation function $\Delta_{ij} \equiv 
\langle \hat{\Theta}_i \hat{\Theta}_j\rangle 
-\langle \hat{\Theta}_i\rangle \langle\hat{\Theta}_j\rangle$ which 
for Hamiltonian systems at criticality decays as $|i-j|^{-\nu}$.
Under the same assumptions used to derive Eqs.~(\ref{Nseq})  and (\ref{newequ}) we can
assume $j$ to be the central site of the MERA (i.e. $j = N/2$).
Suppose then that the associated map $\Phi$ is mixing with fix point $\hat{\rho}_T$.
For any input state $\hat{\rho}_{ij}$ of the sites $i$ and $j$ we then have
$\lim_{\bar{m}\rightarrow \infty} 
( {\cal I}_i\otimes \Phi^{(1\leftarrow \bar{m})}_j )
 ( \hat{\rho}_{ij})
=  \hat{\rho}_i \otimes \hat{\rho}_T$,
with $\hat{\rho}_i \equiv \mbox{Tr}_j [\hat{\rho}_{ij}]$ and ${\cal I}_i$ being the identity 
super-operator of the site $i$.
The speed of convergence, evaluated through the trace
distance, 
is exponentially fast~\cite{TDV,NJP} in $\bar{m}$ and, a part from some constant
pre-factor,  
can be upper-bounded by the quantity $\bar{m}^{d^3}\kappa^{\bar{m}}$ with $\kappa < 1$ 
being the modulus of the largest eigenvalue of $\Phi$
whose associated eigenvector contribute  in the expansion of Eq.~(\ref{faf1}).
This is sufficient for claiming that the distance between 
$( \Phi^{(1\leftarrow \bar{m})}_i\otimes \Phi^{(1\leftarrow \bar{m})}_j )
 ( \hat{\rho}_{ij}^{(\bar{m})}  )$ and 
$\Phi^{(1\leftarrow \bar{m})}_i(\hat{\rho}_i) \otimes\Phi^{(1\leftarrow \bar{m})}_j(\hat{\rho}_j)$
is bounded by $\propto\kappa^{2\bar{m}}$.
Thus we can write 
$
\log_2 \left( |\Delta_{ij} |\right) \leqslant  
2  \bar{m} \log_2 \kappa + {\cal O}(\log_2 \bar{m}),
$
which through the definition of $\bar{m}$
provides a bound for the critical exponent $\nu$ associated to the observable $\hat{\Theta}$ in terms of the properties of the map $\Phi$, i.e. 
$\nu \geqslant- 2 \log_2 \kappa$. In effect one can show that such bound is tight, i.e. 
\begin{equation}
\nu = - 2 \log_2 \kappa\;.
\label{exponent}
\end{equation}
This can be seen for instance by expressing Eq.~(\ref{faf1}) in the Liouville space formalism as in Eq.~(\ref{newequ}), and expanding
the transfer matrices $\hat{E}_{\Phi_i} \otimes \hat{E}_{\Phi_j}$ in Jordan blocks
(the calculation  is similar to the MPS analysis of Ref.~\cite{WOVC}).
{ A numerical test of Eq.~(\ref{exponent})~\cite{NEWPAP}
on  a MERA state approximating the ground energy  
of an Ising chain up to a $10^{-4}$ accuracy 
yielded $\kappa\simeq0.915,0.49,0.52$ to be compared with the exact values 
$\kappa_{th}\simeq0.917,0.46,0.50$
associated with the $x$, $y$ and $z$ two-point correlation functions.
Similar results have been obtained for the XXZ model.}

\paragraph{Concluding  remarks:--}
Equations.~(\ref{TRANSFER}) and~(\ref{exponent}) constitute the main results of our analysis. 
As already discussed by Vidal~\cite{mera}, MERA networks are able to describe algebraic 
decaying correlations. In this work we put on firm grounds this observation giving an explicit 
expression of the critical exponents in terms of properties of the associated QuMERA channels.
The combination of this approach with conformal field theory methods may provide a 
powerful tool to achieve a complete description of one-dimensional critical quantum systems.
Similarly our findings yield a natural connection between the tensor network description 
of the thermodynamic limit of critical systems and the master equation formalism.
Combining these results with the algorithms presented
in~\cite{meraalg, vidallong} one can exploit the introduction of the
transfer operator~(\ref{TRANSFER}) studying directly the infinite size
system improving simulation efficiency. 

The results presented here can be easily extended in several ways. For instance
since a  binary tree  can be seen as a 
MERA with the  disentanglers $\chi$  set to the identity, all 
the arguments presented previously can be easily adapted to this case.
Similarly  also the thermodynamical limit for MPS~\cite{VPC} can be
described in terms of repeated application of CPT maps (here 
the operator~(\ref{TRANSFER}) reduces  to the MPS transfer matrix).
More generally our approach can 
be adapted to any tensor network by associating  it with a family
of CPT transformations which, properly concatenated, allows one to
compute the local observables of the system. 
In this perspective
the quantum circuit~\cite{mera,DEO} associated with the tensor network can  be seen 
as a unitary dilation or, Stinespring representation~\cite{BENGZY}, of the 
corresponding CPT family.
Finally a generalization to higher spatial dimensions seems straightforward.

\acknowledgments We thank M. Rizzi for useful discussions. 
This work was in part founded by the Quantum Information
research program of Centro ``Ennio De Giorgi'' of SNS.


\begin{thebibliography}{99}
\bibitem{NOTA}For easy of notation we consider the dimension $D$ of the tensors $\chi$ and $\lambda$
to be equal to the physical dimension $d$. However, the subsequent equations hold for any $D$.
\bibitem{ReqL} 
  A necessary and sufficient condition~\cite{BENGZY} for
  $\Phi^{(L)} =\Phi^{(R)}$ is the existence 
  of a unitary matrix $U_{{r},{s}}$, such that
  $\hat{L}_{r} = \sum_{s} U_{{r},{s}} \; \hat{R}_{{s}}$. This can then
  be easily casted   into a necessary and sufficient
  condition for the tensor ${\cal M}_5$.
\end{thebibliography}
\end{document}